\documentclass[10pt,twocolumn,letterpaper]{article}
\usepackage[top=0.75in,bottom=0.8in,left=0.62in,right=0.62in,columnsep=0.25in]{geometry}
\usepackage{mathptmx}
\usepackage{amsmath,amssymb,bm}
\usepackage{graphicx,booktabs}
\usepackage{tikz}
\usetikzlibrary{decorations.pathmorphing,decorations.markings,arrows.meta}
\usepackage[font=small,labelfont=bf,justification=raggedright,singlelinecheck=false]{caption}
\usepackage{titlesec}
\usepackage{xcolor}
\titleformat{\section}{\normalfont\small\bfseries\centering}{\thesection.}{0.5em}{\MakeUppercase}
\titleformat{\subsection}{\normalfont\small\bfseries\centering}{\Alph{subsection}.}{0.5em}{}
\titlespacing*{\section}{0pt}{1.4ex}{0.8ex}
\titlespacing*{\subsection}{0pt}{1.2ex}{0.6ex}
\renewcommand{\thesection}{\Roman{section}}

\newcommand{\bk}{\mathbf{k}}
\newcommand{\bp}{\mathbf{p}}
\newcommand{\bq}{\mathbf{q}}
\newcommand{\sgn}{\operatorname{sgn}}
\tikzset{
 fermion/.style={thick,postaction={decorate},decoration={markings,mark=at position .55 with {\arrow{Stealth[length=4pt]}}}},
 coul/.style={decorate,decoration={snake,amplitude=1.1pt,segment length=4pt},thin},
 vtx/.style={circle,fill=black,inner sep=1.1pt}
}
\begin{document}
\twocolumn[{
\begin{center}
{\large\bfseries Exchange self-energy and vertex corrections to static screening in the\\ two-dimensional electron gas}\\[8pt]
{\normalsize Sankar Das Sarma}\\[3pt]
{\small\itshape Condensed Matter Theory Center and Joint Quantum Institute, Department of Physics, University of Maryland, College Park, Maryland 20742, USA}\\[10pt]
\begin{minipage}{0.84\textwidth}
\small
We compute analytically the leading beyond-ring corrections to the static irreducible polarizability of the two-dimensional (2D) electron gas, namely the two exchange self-energy diagrams and the exchange vertex diagram, each of first order in the Coulomb coupling, and insert the resulting polarizability into the RPA geometric series to obtain the static dielectric function and screened interaction. The self-energy and vertex contributions individually diverge as $\ln(k_F/q)$ for $q\to0$; we obtain both divergent pieces in closed form, including their constant terms, $\mp(r_s/\sqrt2\pi)[\ln(k_F/q)+4\ln2+3/2]$ in units of the density of states. Their sum is finite and reproduces exactly the Hartree--Fock compressibility, which renormalizes the Thomas--Fermi wave vector to $q_{TF}(1+\sqrt2 r_s/\pi)$. The linear-in-$q$ correction vanishes identically. The leading wave-vector dependence is nonanalytic, $\propto r_s(q/k_F)^2[\ln(q/k_F)+\beta]$: the coefficient of the logarithm is obtained analytically and $\beta=6.6441(1)$ numerically. Equivalently, the exchange local-field factor is $G_x(q)=(q/\pi k_F)\{1+(q/k_F)^2[\ln(q/k_F)+\beta]/96\}$. We then compare critically with the Hubbard local-field factor $G_H=q/2\sqrt{q^2+k_F^2}$. $G_H$ overestimates the exact long-wavelength slope by a factor $\pi/2$, which causes a violation of the compressibility sum rule already at $O(r_s)$. It underestimates $G_x$ by a factor $\approx3$ near $2k_F$, and it lacks the nonanalytic structure. Diffusion Monte Carlo compressibilities confirm that the exchange slope $1/\pi$ remains within 4\% (8\%) of the exact value at $r_s=1$ ($r_s=2$), whereas the Hubbard slope is off by about 50\%. Extending the calculation to $q=4k_F$, we show that exchange converts the one-sided 2D Kohn kink at $2k_F$ into a sharp two-sided peak, a qualitative (though at $r_s\gtrsim1$ not quantitatively controlled) improvement over Thomas--Fermi and RPA screening. In real space this makes the Friedel oscillations around a charged impurity 2--3 times larger than in RPA at $r_s=0.5$--1 for $k_Fr\lesssim50$, with an amplitude that increases with distance over the range studied; implications for Kohn--Luttinger pairing are discussed. Relation to earlier numerical work on the same diagrams is discussed. All results are expressed in terms of $r_s$ and $q/k_F$.
\end{minipage}
\end{center}
\vspace{10pt}
}]

\section{Introduction}
The random-phase approximation (RPA) for the static dielectric function of the 2D electron gas uses the noninteracting bubble as the irreducible polarizability. In 2D this bubble is strictly $q$-independent for $q\le2k_F$, so RPA reduces to Thomas--Fermi (TF) screening, $\varepsilon=1+q_{TF}/q$, over the whole range $q\le 2k_F$. The first corrections beyond the ring approximation come from the three diagrams of first order in the Coulomb interaction: two exchange self-energy insertions and one exchange vertex correction [Fig.~\ref{fig:diagrams}(b)--(d)].

This note evaluates these diagrams in the static limit. The dressed polarizability is inserted into the standard RPA series (Fig.~\ref{fig:series}), and the correction to TF screening is extracted to the leading two orders in $q$. The first-order static polarizability of the 2D gas has been computed numerically before~\cite{Maldague78,Czachor82,Khalil01}; our focus is its analytic long-wavelength structure (see Sec.~\ref{sec:prior}).

A widely used alternative is the Hubbard approximation~\cite{Hubbard57,Jonson76}. It sums the exchange ladder crudely, with the Coulomb interaction replaced by its value at a typical momentum $\sim k_F$, and omits the self-energy insertions. In Sec.~\ref{sec:hubbard} we compare the two local-field factors critically.

Two features of the 2D problem make this nontrivial. First, the self-energy and vertex diagrams are separately logarithmically divergent as $q\to0$, reflecting the logarithmic divergence of the Hartree--Fock Fermi velocity. Second, the first $q$ dependence is nonanalytic.

\section{Model and RPA}
We use $\hslash=1$ and a background dielectric constant $\kappa$. The basic quantities are
\begin{gather}
v_q=\frac{2\pi e^2}{\kappa q},\qquad a_B=\frac{\kappa}{me^2},\qquad r_s=\frac{\sqrt2}{k_Fa_B},\nonumber\\
N_F=\frac{m}{\pi},\qquad q_{TF}=v_qN_F\,q=\frac{2}{a_B}=\sqrt2\,r_sk_F,\label{eq:units}
\end{gather}
where $n=k_F^2/2\pi$ (spin-degenerate) and $x\equiv q/k_F$. The Coulomb scale that appears throughout is
\begin{equation}
\frac{m^2e^2}{\kappa k_F}=N_F\,\frac{\pi}{k_Fa_B}=N_F\,\frac{\pi r_s}{\sqrt2}.\label{eq:scale}
\end{equation}

With $\Pi(q)$ the irreducible (proper) static polarizability, the RPA geometric series gives
\begin{equation}
\chi(q)=\frac{\Pi(q)}{1-v_q\Pi(q)},\qquad \varepsilon(q)=1-v_q\,\Pi(q).\label{eq:series}
\end{equation}
The static Lindhard function in 2D is
\begin{align}
\Pi_0(q)&=2\!\int\!\!\frac{d^2k}{(2\pi)^2}\,\frac{f_{\bk}-f_{\bk+\bq}}{\xi_{\bk}-\xi_{\bk+\bq}}\nonumber\\
&=-N_F\Big[1-\Theta(x-2)\sqrt{1-4/x^2}\,\Big],\label{eq:Pi0}
\end{align}
so that $\varepsilon_{\rm RPA}=1+q_{TF}/q$ exactly for $q\le 2k_F$.

\newcommand{\bub}[3]{
 \begin{scope}[xshift=#1cm]
 \draw[fermion] (0,0) arc[start angle=180,end angle=0,x radius=#2cm,y radius=#3cm];
 \draw[fermion] ({2*#2},0) arc[start angle=0,end angle=-180,x radius=#2cm,y radius=#3cm];
 \node[vtx] at (0,0) {}; \node[vtx] at ({2*#2},0) {};
 \end{scope}}
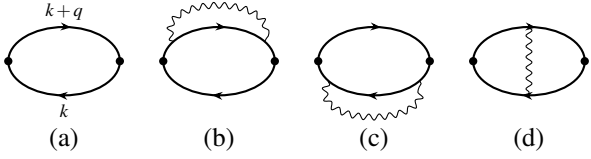
\begin{figure}[t]
\centering
\begin{tikzpicture}[scale=0.82]
\bub{0}{0.9}{0.55}
\node at (0.9,0.8) {\scriptsize $k+q$}; \node at (0.9,-0.8) {\scriptsize $k$};
\node at (0.9,-1.25) {(a)};
\bub{2.5}{0.9}{0.55}
\draw[coul] (2.663,0.315) to[out=115,in=65,looseness=1.4] (4.137,0.315);
\node at (3.4,-1.25) {(b)};
\bub{5.0}{0.9}{0.55}
\draw[coul] (5.163,-0.315) to[out=-115,in=-65,looseness=1.4] (6.637,-0.315);
\node at (5.9,-1.25) {(c)};
\bub{7.5}{0.9}{0.55}
\draw[coul] (8.4,0.55) -- (8.4,-0.55);
\node at (8.4,-1.25) {(d)};
\end{tikzpicture}
\caption{Irreducible static polarizability to first order in $e^2$. (a) Ring (bubble) $\Pi_0$. (b), (c) Exchange self-energy insertions on the particle and hole lines, $\Pi_{\rm SE}$. (d) Exchange vertex correction, $\Pi_{\rm V}$. Wavy lines denote the instantaneous Coulomb interaction $v_{\bk-\bk'}$. The Hartree tadpole is cancelled by the neutralizing background.}
\label{fig:diagrams}
\end{figure}

\begin{figure}[t]
\centering
\begin{tikzpicture}[scale=0.8]
\def\blob#1{\draw[thick,fill=gray!30] (#1,0) ellipse (0.4 and 0.28); \node at (#1,0) {\scriptsize $\Pi$};}
\node at (-0.35,0) {$\chi\;=$};
\blob{0.6}
\node at (1.35,0) {$+$};
\blob{2.1} \draw[coul] (2.5,0)--(3.2,0); \blob{3.6}
\node at (4.35,0) {$+$};
\blob{5.1} \draw[coul] (5.5,0)--(6.2,0); \blob{6.6} \draw[coul] (7.0,0)--(7.7,0); \blob{8.1}
\node at (9.0,0) {$+\,\cdots$};
\end{tikzpicture}\\[8pt]
\begin{tikzpicture}[scale=0.8]
\node at (-0.65,0) {$\Pi\;=$};
\bub{0}{0.55}{0.38}
\node at (1.45,0) {$+$};
\bub{1.8}{0.55}{0.38}
\draw[coul] (1.899,0.218) to[out=115,in=65,looseness=1.4] (2.801,0.218);
\node at (3.25,0) {$+$};
\bub{3.6}{0.55}{0.38}
\draw[coul] (3.699,-0.218) to[out=-115,in=-65,looseness=1.4] (4.601,-0.218);
\node at (5.05,0) {$+$};
\bub{5.4}{0.55}{0.38}
\draw[coul] (5.95,0.38) -- (5.95,-0.38);
\end{tikzpicture}
\caption{RPA geometric series built from the irreducible polarizability $\Pi=\Pi_0+\Pi_{\rm SE}+\Pi_{\rm V}$ of Fig.~\ref{fig:diagrams}. This gives Eq.~(\ref{eq:series}).}
\label{fig:series}
\end{figure}

\section{First-order diagrams}
\subsection{Matsubara sums}
Let $G(\bk,i\omega_n)=(i\omega_n-\xi_{\bk})^{-1}$ and $D_{\bk}\equiv\xi_{\bk}-\xi_{\bk+\bq}$. At zero external frequency, and with a frequency-independent (instantaneous) interaction, every frequency sum factorizes. The two identities we need are
\begin{align}
L_{\bk}&\equiv T\sum_n G(\bk,i\omega_n)G(\bk+\bq,i\omega_n)=\frac{f_{\bk}-f_{\bk+\bq}}{D_{\bk}},\label{eq:L}\\
T\sum_n G^2(\bk,i\omega_n)&G(\bk+\bq,i\omega_n)=\frac{\partial L_{\bk}}{\partial\xi_{\bk}}.\label{eq:dL}
\end{align}
Note that $L_{\bk}\le0$. The static exchange self-energy is
\begin{equation}
\Sigma_x(\bk)=-\!\int\!\!\frac{d^2k'}{(2\pi)^2}\,v_{\bk-\bk'}f_{\bk'}.\label{eq:Sigx}
\end{equation}
Its closed form is given in Appendix~\ref{app:sigma}. It satisfies $\Sigma_x(k_F)=-2e^2k_F/\pi\kappa$, and its slope diverges logarithmically at the Fermi surface, $\partial_k\Sigma_x\simeq(e^2/\pi\kappa)\ln(8k_F/|k-k_F|)$.

\subsection{Self-energy diagrams}
By Eq.~(\ref{eq:dL}), diagrams (b)+(c) equal the first-order change of the Lindhard function under $\xi_{\bk}\to\xi_{\bk}+\Sigma_x(\bk)$.

We work at fixed density, i.e., with the counterterm $\delta\mu=\Sigma_x(k_F)$. Because $\Sigma_x$ is isotropic and the HF dispersion is monotonic, the Fermi surface is unchanged, and the terms $\propto\delta(\xi)$ generated by differentiating $f$ vanish. (They are in any case $\propto\partial\Pi_0/\partial\mu=0$ in 2D.) Hence
\begin{equation}
\Pi_{\rm SE}(q)=-2\!\int_{\bk}\frac{(f_{\bk}-f_{\bk+\bq})\,[\Sigma_x(\bk)-\Sigma_x(\bk+\bq)]}{D_{\bk}^2},\label{eq:PiSE}
\end{equation}
with $\int_{\bk}\equiv\int d^2k/(2\pi)^2$.

\subsection{Vertex diagram}
Diagram (d) is the first iterate of the static exchange ladder for the density vertex,
\[
\Lambda_{\bk}=1-\int_{\bk'}v_{\bk-\bk'}L_{\bk'}\Lambda_{\bk'},\qquad \Pi=2\int_{\bk}L_{\bk}\Lambda_{\bk}.
\]
This gives
\begin{equation}
\Pi_{\rm V}(q)=-2\!\int_{\bk}\!\int_{\bk'}L_{\bk}\,v_{\bk-\bk'}\,L_{\bk'}.\label{eq:PiV}
\end{equation}
The sign is such that exchange enhances $|\Pi|$. It is checked independently by the compressibility sum rule in Sec.~\ref{sec:q0}.

\subsection{A convergent combined kernel}
Shifting $\bk'\to\bk'+\bq$ in one of the two terms gives the exact identity
\begin{equation}
\Sigma_x(\bk)-\Sigma_x(\bk+\bq)=-\!\int_{\bk'}v_{\bk-\bk'}\,(f_{\bk'}-f_{\bk'+\bq}).\label{eq:Sigdiff}
\end{equation}
Inserting Eq.~(\ref{eq:Sigdiff}) into Eq.~(\ref{eq:PiSE}) yields
\[
\Pi_{\rm SE}=2\int_{\bk}\int_{\bk'}v_{\bk-\bk'}\,L_{\bk}L_{\bk'}\,\frac{D_{\bk'}}{D_{\bk}}.
\]
Symmetrizing in $\bk\leftrightarrow\bk'$ and adding Eq.~(\ref{eq:PiV}) gives
\begin{equation}
\boxed{\Pi_1\equiv\Pi_{\rm SE}+\Pi_{\rm V}=\!\int_{\bk}\!\int_{\bk'}\!v_{\bk-\bk'}\,\Delta f_{\bk}\Delta f_{\bk'}\Big(\frac{1}{D_{\bk}}-\frac{1}{D_{\bk'}}\Big)^{2}}\label{eq:kernel}
\end{equation}
with $\Delta f_{\bk}=f_{\bk}-f_{\bk+\bq}$.

Equation~(\ref{eq:kernel}) is the 2D counterpart of the symmetrized first-order exchange kernel used in 3D~\cite{EngelVosko90}. Its important property is that the squared bracket vanishes as $\bk'\to\bk$. This removes the $v_{\bk-\bk'}$ singularity responsible for the separate logarithmic divergences of $\Pi_{\rm SE}$ and $\Pi_{\rm V}$.

\section{Exact reduction for $q<2k_F$}
\subsection{Lune geometry}
Define $\bp=\bk+\bq/2$ and take $\bq=q\hat{x}$. Then
\[
D_{\bk}=-\frac{q\,p_x}{m},\qquad \Delta f_{\bk}=\sgn(p_x)\,\chi_{\mathcal L}(\bp),
\]
where $\mathcal L$ is the symmetric difference of the two Fermi disks centered at $\mp\bq/2$ (two ``lunes''). Solving $|\bp\mp\bq/2|=k_F$ at polar angle $\theta$, each lune is parametrized exactly by
\begin{gather}
p=r(\cos\theta,\sin\theta),\qquad r=\rho(\theta)+q|\cos\theta|\,t,\nonumber\\
\rho(\theta)=\sqrt{k_F^2-\tfrac14q^2\sin^2\theta},\qquad t\in[-\tfrac12,\tfrac12].\label{eq:lune}
\end{gather}
The radial width is exactly $q|\cos\theta|$. The Jacobian therefore takes the uniform form
\begin{equation}
\frac{d^2p}{|p_x|}=\frac{r\,dr\,d\theta}{r|\cos\theta|}=q\,d\theta\,dt.\label{eq:jac}
\end{equation}

\subsection{Reduced integral}
Using Eqs.~(\ref{eq:lune})--(\ref{eq:jac}) in Eq.~(\ref{eq:kernel}), with all momenta measured in units of $k_F$, gives
\begin{gather}
\Pi_1(q)=\frac{m^2e^2}{\kappa k_F}\,P(x),\qquad P(x)=\frac{I(x)}{8\pi^3},\nonumber\\
I(x)=\int_0^{2\pi}\!\!d\theta\,d\theta'\!\int_{-1/2}^{1/2}\!\!dt\,dt'\;
\frac{(p_x-p'_x)^2}{p_x\,p'_x\,|\bp-\bp'|}.\label{eq:I}
\end{gather}
The same steps applied to the two pieces separately give
\begin{align}
P_{\rm V}&=-\frac{1}{4\pi^3}\int d\theta\,dt\,d\theta'\,dt'\;\frac{1}{|\bp-\bp'|},\label{eq:PV}\\
P_{\rm SE}&=\frac{1}{2\pi^2x}\int d\theta\,dt\;\frac{\Sigma_x(|\bp+\tfrac{\bq}{2}|)-\Sigma_x(|\bp-\tfrac{\bq}{2}|)}{r\cos\theta},\label{eq:PSE}
\end{align}
where $\Sigma_x$ is in units of $e^2k_F/\kappa$.

An equivalent unsymmetrized form, well suited to numerics, is
\[
P=\frac{1}{4\pi^3}\int d\theta\,dt\,\frac{1}{p_x}\int d\theta'\,dt'\,\frac{p'_x-p_x}{|\bp-\bp'|}.
\]
Here the inner integral is odd under $p_x\to-p_x$, since $\mathcal L$ is mirror-symmetric. It therefore cancels the $1/p_x$ at the lune tips.

\subsection{Two exact identities}
Let $\phi=\theta-\theta'$ and $\lambda=\ln(r/r')$. Then
\begin{align}
|\bp-\bp'|^2&=4rr'\big[\sin^2(\phi/2)+\sinh^2(\lambda/2)\big],\label{eq:dist}\\
\frac{(p_x-p'_x)^2}{p_xp'_x}&=4\sinh^2\!\Big(\frac12\ln\frac{p_x}{p'_x}\Big)\quad(p_xp'_x>0).\label{eq:num}
\end{align}
For $p_xp'_x<0$ the left side of Eq.~(\ref{eq:num}) is negative and equals $-4\cosh^2[\frac12\ln|p_x/p'_x|]$. These identities make the structure of the small-$x$ expansion transparent.

\begin{figure*}[t]
\centering
\includegraphics[width=0.86\textwidth]{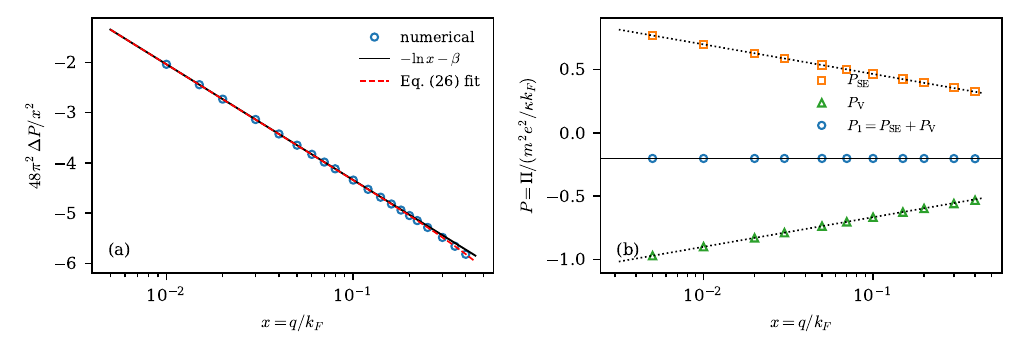}
\caption{(a) $48\pi^2\Delta P/x^2$ versus $x$ (circles), compared with the asymptotic form $-\ln x-\beta$, $\beta=6.6441$ (solid), and with the subleading fit of Eq.~(\ref{eq:fit}) (dashed). (b) Self-energy (squares), vertex (triangles), and total (circles) first-order polarizability. Dotted curves: the analytic asymptotes of Eqs.~(\ref{eq:PVas})--(\ref{eq:PSEas}). Horizontal line: $P(0)=-2/\pi^2$.}
\label{fig:num}
\end{figure*}

\section{The limit $q\to0$ and the compressibility}\label{sec:q0}
\subsection{Finite total}
At $x=0$ we have $r=r'=1$ and $|\bp-\bp'|=2|\sin(\phi/2)|\equiv g(\phi)^{-1}$. For even $g$, rotational invariance gives $\int d\theta'\cos\theta'\,g(\theta-\theta')=\cos\theta\int d\phi\cos\phi\,g(\phi)$. Expanding $(c-c')^2/cc'=c/c'+c'/c-2$ then yields
\begin{align}
I(0)&=4\pi\!\int_0^{2\pi}\!\!d\phi\,(\cos\phi-1)g(\phi)=-4\pi\!\int_0^{2\pi}\!\!d\phi\,\sin\frac{\phi}{2}\nonumber\\
&=-16\pi.\label{eq:I0}
\end{align}
Hence $P(0)=-2/\pi^2$, and by Eq.~(\ref{eq:scale})
\begin{equation}
\Pi(q\to0)=-N_F\Big(1+\frac{\sqrt2}{\pi}r_s\Big).\label{eq:comp}
\end{equation}

\subsection{Consistency checks}
Equation~(\ref{eq:comp}) is exactly the HF compressibility sum rule. The HF energy per particle in 2D is $\epsilon_x=-4e^2k_F/3\pi\kappa$. This gives $\mu_x=-2e^2k_F/\pi\kappa$ and $d\mu/dn=N_F^{-1}(1-\sqrt2 r_s/\pi)$, i.e., $\kappa_0/\kappa=1-\sqrt2r_s/\pi$.

In Landau language, $\Pi(0)=-N_F(m^*/m)/(1+F_0^s)$ with $f^s_{\bk\bk'}=-v_{\bk-\bk'}/2$. The combination $F_1^s-F_0^s=(N_F/2)\langle v(1-\cos\phi)\rangle=\sqrt2r_s/\pi$ is finite, while $F_0^s$ and $m^*/m-1=F_1^s$ are separately log-divergent.

\subsection{Separate self-energy and vertex pieces}
In Eq.~(\ref{eq:PV}), for small $x$, $\int d\theta'\,|\bp-\bp'|^{-1}\simeq\int_{-\pi}^{\pi}d\phi\,[4\sin^2(\phi/2)+\lambda^2]^{-1/2}=2\ln(8/|\lambda|)+O(\lambda^2)$. Here $\lambda\simeq x|\cos\theta|(t-t')$. Averaging with $\langle\ln|t-t'|\rangle=-3/2$ and $\langle\ln|\cos\theta|\rangle_\theta=-\ln2$ gives
\begin{align}
\Pi_{\rm V}&=-N_F\frac{r_s}{\sqrt2\,\pi}\Big[\ln\frac{k_F}{q}+4\ln2+\frac32\Big]+O(x^2\ln x),\label{eq:PVas}\\
\Pi_{\rm SE}&=+N_F\frac{r_s}{\sqrt2\,\pi}\Big[\ln\frac{k_F}{q}+4\ln2-\frac12\Big]+O(x^2\ln x).\label{eq:PSEas}
\end{align}
The leading term of Eq.~(\ref{eq:PSEas}) is $N_F\,\partial_k\Sigma_x/v_F$ evaluated at $|k-k_F|\sim q$, i.e., the vanishing HF density of states. Both expansions are confirmed numerically to $10^{-4}$ relative accuracy [Fig.~\ref{fig:num}(b)].

\section{Small-$q$ expansion}
\subsection{Absence of a linear term}
By Eq.~(\ref{eq:lune}), $\ln r=x|\cos\theta|\,t+O(x^2)$ is odd in $t$. Away from the diagonal $\phi=0$, the integrand of Eq.~(\ref{eq:I}) is a smooth function of $(\ln r,\ln r')$. Its $O(x)$ term is therefore odd under $(t,t')\to(-t,-t')$ and integrates to zero.

The only nonanalyticity sits on the diagonal. There, by Eqs.~(\ref{eq:dist})--(\ref{eq:num}), the kernel behaves as $|\lambda|$, but only in a strip $|\phi|\lesssim|\lambda|\sim x$. This strip contributes at $O(x^2)$. Hence
\[
I(x)=-16\pi+O(x^2\ln x).
\]
A free fit of the numerical data (Sec.~\ref{sec:num}) gives a linear coefficient $|a_1|<3\times10^{-7}$, compared with $P(0)\approx0.2$.

\subsection{Coefficient of the logarithm}
\emph{Outer region} ($x\ll|\phi|$). Expand Eqs.~(\ref{eq:dist})--(\ref{eq:num}) to second order in $\lambda$ and in $\sigma=\frac12\ln rr'$. The required $t$-averages are
\[
\langle\lambda^2\rangle=\frac{x^2(c^2+c'^2)}{12},\quad \langle\lambda_2\rangle=\frac{x^2(c^2-c'^2)}{12},\]\[\langle\sigma_1^2\rangle=\frac{x^2(c^2+c'^2)}{48},
\]
and so on, with $c=\cos\theta$ and $s=\sin\theta$. The only terms that are non-integrable at $\phi\to0$ are
\begin{equation}
K^{(2)}\simeq\frac{x^2\,(2c^2-s^2)}{12\,|\phi|}.\label{eq:outer}
\end{equation}
They come from the $\lambda^2$ term of the numerator and from the $\sinh^2(\lambda/2)$ term of the denominator.

\emph{Inner region} ($|\phi|\sim x|c|$). Here $\ln(p_x/p'_x)\simeq\lambda-\phi\tan\theta$, and the kernel becomes
\[
K\simeq\frac{(\lambda-\phi\tan\theta)^2}{\sqrt{\phi^2+\lambda^2}}.
\]
Integrating over $|\phi|<\Phi$ gives
\[
\lambda^2(2-\tan^2\theta)\ln\frac{2\Phi}{|\lambda|}+\tan^2\theta\,\Phi^2+O(\lambda^2).
\]
This matches Eq.~(\ref{eq:outer}) and cuts the logarithm off at $|\phi|\sim x|c|$. Integrating over $\theta$, using $\int_0^{2\pi}(2c^2-s^2)d\theta=\pi$, gives
\begin{equation}
I(x)=-16\pi+\frac{\pi}{6}\,x^2\ln\frac1x+C\,x^2+O(x^4\ln x).\label{eq:Iexp}
\end{equation}
The constant $C$ collects the finite parts of the inner and outer regions, including $\langle\ln|t-t'|\rangle$, $\langle\ln|\cos\theta|\rangle$, and Hadamard finite parts of Eq.~(\ref{eq:outer}). We determine it numerically.

\subsection{Result}
Writing $C=-(\pi/6)\beta$, we have
\begin{equation}
P(x)=-\frac{2}{\pi^2}-\frac{x^2}{48\pi^2}\big[\ln x+\beta\big]+O(x^4\ln x),\label{eq:Pres}
\end{equation}
with
\[
\beta=6.6441(1).
\]

\section{Numerical evaluation}\label{sec:num}
\subsection{Method}
The radial ($t'$) integrations in Eqs.~(\ref{eq:I}) and (\ref{eq:PV}) are elementary. With $R=|\bp-\bp'|$, the needed primitives are $\int dr'/R=\operatorname{arcsinh}[(r'-r\cos\phi)/r|\sin\phi|]$ and $\int r'dr'/R=R+r\cos\phi\int dr'/R$.

The remaining three-dimensional integrals were done by composite Gauss--Legendre quadrature. The $\theta'$ panels are geometrically graded toward $\theta'=\theta$, down to $10^{-10}$, and toward the lune tips. $P_{\rm SE}$ was computed independently from Eq.~(\ref{eq:PSE}) using the exact $\Sigma_x$ of Eq.~(\ref{eq:SigA}).

\subsection{Checks}
Three independent checks agree:
\begin{itemize}
\item $P_{\rm SE}+P_{\rm V}$ agrees with the combined form to $\sim10^{-11}$.
\item $P(x\to0)\to-2/\pi^2$.
\item A free fit to $\Delta P\equiv P(x)-P(0)$ over $0.01\le x\le0.4$ with basis $\{x^2,x^2\ln x,x^4,x^4\ln x,\dots\}$ returns the $x^2\ln x$ coefficient $-1.00002/(48\pi^2)$, versus the analytic $-1/(48\pi^2)$.
\end{itemize}

\subsection{Fit for $\beta$}
Fixing the logarithm coefficient at its analytic value, fits with 3--5 parameters over three different $x$ windows give $\beta=6.64407$--$6.64409$. Including subleading terms, the fit is
\begin{equation}
\frac{48\pi^2\,\Delta P}{x^2}\simeq-\big[\ln x+\beta+0.81\,x^2+0.28\,x^2\ln x\big].\label{eq:fit}
\end{equation}
The subleading coefficients are fit-dependent at the $\sim\!10\%$ level. Representative values are given in Table~\ref{tab:num} and Fig.~\ref{fig:num}. Changing the Gauss--Legendre orders (12$\to$20 nodes per panel in $t$) changes $P(x)$ by $<10^{-10}$ for $x\le0.3$, by $\sim10^{-6}$ at $x=1.9$, and by $\lesssim10^{-4}$ within $10^{-3}k_F$ of $2k_F$. The quoted uncertainty of $\beta$ is the spread over fit windows and basis sets. The numerical code and data are available from the author upon request.

\begin{table}[t]
\caption{Numerical values of the self-energy, vertex, and total first-order polarizability in units of $m^2e^2/\kappa k_F$ [Eq.~(\ref{eq:scale})], and $\Delta P=P(x)+2/\pi^2$.}
\label{tab:num}
\centering\small
\begin{tabular}{@{}lcccc@{}}
\toprule
$x$ & $P_{\rm SE}$ & $P_{\rm V}$ & $P_1$ & $\Delta P$\\
\midrule
0.005 & 0.767096 & $-0.969738$ & $-0.202642$ & $-7.09\times10^{-8}$\\
0.01 & 0.696870 & $-0.899513$ & $-0.202643$ & $-4.303\times10^{-7}$\\
0.03 & 0.585599 & $-0.788248$ & $-0.202648$ & $-5.960\times10^{-6}$\\
0.05 & 0.533914 & $-0.736576$ & $-0.202662$ & $-1.925\times10^{-5}$\\
0.1 & 0.463971 & $-0.666706$ & $-0.202734$ & $-9.168\times10^{-5}$\\
0.2 & 0.394666 & $-0.597735$ & $-0.203069$ & $-4.263\times10^{-4}$\\
0.3 & 0.354865 & $-0.558549$ & $-0.203684$ & $-1.0418\times10^{-3}$\\
0.4 & 0.327279 & $-0.531887$ & $-0.204608$ & $-1.9655\times10^{-3}$\\
\bottomrule
\end{tabular}
\end{table}

\section{Screening}
\subsection{Irreducible polarizability}
Combining Eqs.~(\ref{eq:Pi0}), (\ref{eq:scale}), and (\ref{eq:Pres}), for $x\ll1$,
\begin{equation}
\boxed{\;\Pi(q)=-N_F\Big\{1+\frac{\sqrt2}{\pi}r_s+\frac{r_s}{48\sqrt2\,\pi}\,x^2\big[\ln x+\beta\big]\Big\}\;}\label{eq:Pifinal}
\end{equation}
up to $O(r_s\,x^4\ln x)$ and $O(r_s^2)$ corrections.

\subsection{Local-field factor}
Defining $\Pi=\Pi_0/(1+v_qG\Pi_0)$, Eq.~(\ref{eq:Pifinal}) is equivalent to the exchange local-field factor
\begin{equation}
G_x(q)=\frac{x}{\pi}\Big\{1+\frac{x^2}{96}\big[\ln x+\beta\big]\Big\},\label{eq:G}
\end{equation}
which is independent of $r_s$, as it must be at this order. The leading term $x/\pi$ is fixed by the compressibility sum rule.

\subsection{Dielectric function and screened interaction}
The RPA series, Eq.~(\ref{eq:series}), then gives
\begin{align}
\varepsilon(q)&=1+\frac{q^*_{TF}}{q}+\frac{r_s^2}{48\pi}\,\frac{q}{k_F}\Big[\ln\frac{q}{k_F}+\beta\Big],\label{eq:eps}\\
q^*_{TF}&=q_{TF}\Big(1+\frac{\sqrt2}{\pi}r_s\Big)=\sqrt2\,r_sk_F\Big(1+\frac{\sqrt2}{\pi}r_s\Big),\nonumber
\end{align}
and
\begin{equation}
W(q)=\frac{v_q}{\varepsilon(q)}=\frac{2\pi e^2/\kappa}{\,q+q^*_{TF}+\dfrac{r_s^2k_F}{48\pi}\,x^2\big[\ln x+\beta\big]}.\label{eq:W}
\end{equation}

\subsection{Summary of the TF corrections}
There are two leading corrections to TF screening:
\begin{enumerate}
\item The inverse screening length is renormalized, $q_{TF}\to q^*_{TF}$. This is an $O(r_s)$ relative enhancement. The Hubbard-type resummation $\Pi_0/(1+v_qG_x\Pi_0)$ gives instead $q_{TF}/(1-\sqrt2r_s/\pi)$, which is identical at this order.
\item The first wave-vector dependence of $q\,\varepsilon(q)-q$ is of order $x^2\ln x$, not $x$. This contrasts with 3D, where exchange gives $G_x\propto x^2$.
\end{enumerate}

The bracket $\ln x+\beta$ changes sign at $x_0=e^{-\beta}\approx1.3\times10^{-3}$. For $x>x_0$ the $q$-dependent term further enhances screening.

\subsection{Caveats}
Equation~(\ref{eq:eps}) is complete to first order in $e^2$ in $\Pi$. The $O(r_s^2)$ correction to the $q$-independent part of $q^*_{TF}$ requires second-order (two-interaction-line) diagrams, including correlation, and is not included. This is the natural next step. At the same order, the exact compressibility (e.g., from quantum Monte Carlo) can be used to fix $q^*_{TF}$ nonperturbatively.

\section{Is this an improvement on Thomas--Fermi screening? The $2k_F$ Kohn anomaly}\label{sec:kohn}
\subsection{Extending the calculation beyond $2k_F$}
To address this we extended the numerical evaluation of Eq.~(\ref{eq:kernel}) to $q>2k_F$. There the two Fermi disks centered at $\pm\bq/2$ no longer overlap. Each lune becomes a full disk, parametrized by $r=\frac q2|\cos\theta|+2\rho(\theta)\,t$ for $|\sin\theta|<2k_F/q$, with measure $d^2p/|p_x|=dr\,d\theta/|\cos\theta|$.

The code reproduces the exact $\Pi_0$ to machine precision. An independent evaluation of $\Pi_{\rm SE}+\Pi_{\rm V}$ agrees with the combined kernel to $\sim10^{-7}$.

\begin{figure*}[t]
\centering
\includegraphics[width=0.86\textwidth]{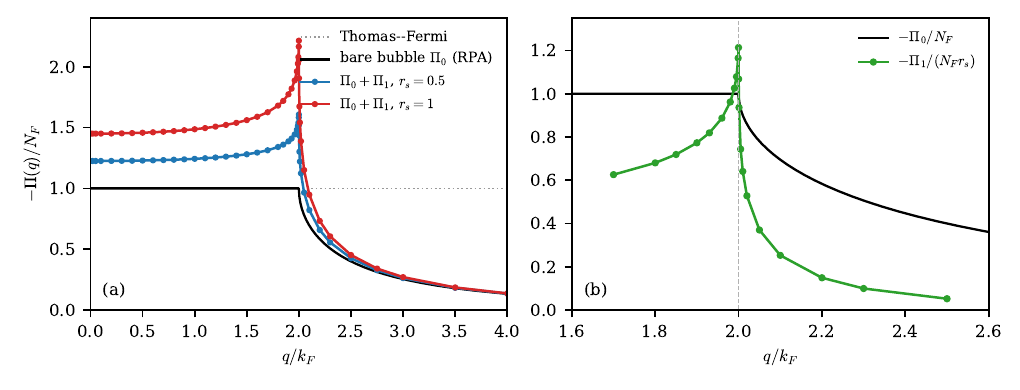}
\caption{(a) Static irreducible polarizability $-\Pi/N_F$ from $0$ to $4k_F$. Solid black: bare bubble $\Pi_0$ (RPA), Eq.~(\ref{eq:Pi0}). Dotted: Thomas--Fermi, $-\Pi/N_F\equiv1$. Colored: $\Pi_0+\Pi_1$ at $r_s=0.5$ and $1$. (b) Zoom near $2k_F$: bare bubble versus the first-order correction per unit $r_s$, $-\Pi_1/(N_Fr_s)=-(\pi/\sqrt2)P(x)$. The bare Kohn anomaly is a one-sided square-root kink. The exchange correction develops a sharp peak at $2k_F$, with singular slopes on both sides. Symbols are computed points; lines are guides to the eye.}
\label{fig:kohn}
\end{figure*}

Figure~\ref{fig:kohn}(a) shows $-\Pi/N_F$ for $0\le q\le4k_F$, and Fig.~\ref{fig:kohn}(b) zooms onto $2k_F$.

\subsection{Long wavelengths ($q\ll k_F$)}
Here the answer is an unambiguous, controlled improvement. TF (and hence RPA) screening uses the noninteracting compressibility. The present result replaces $q_{TF}$ by $q_{TF}^*=q_{TF}(1+\sqrt2r_s/\pi)$, which is exact to $O(r_s)$. The resummed form $q_{TF}/(1-\sqrt2r_s/\pi)$ reproduces the compressibility through $O(r_s^2)$ (Sec.~\ref{sec:hubbard}). At $r_s=1$ this is a 45\% increase of the inverse screening length.

The accompanying $q$ dependence ($\propto x^2\ln x$) is weak: $P$ changes by only 8\% between $q=0$ and $q=k_F$. For $q\lesssim k_F$ the improvement over TF is therefore essentially a renormalization of the screening length.

\subsection{The $2k_F$ region}
TF theory has no $2k_F$ physics at all: it continues $-\Pi/N_F=1$ to all $q$. The 2D RPA bubble is identical to TF for $q\le2k_F$~\cite{Stern67}. Its only memory of the Fermi surface is the Kohn anomaly, a one-sided kink: $\Pi_0$ is flat below $2k_F$ and falls as $\sqrt{1-4k_F^2/q^2}$ above, so its derivative diverges only as $q\to2k_F^+$.

The exchange diagrams change this qualitatively [Fig.~\ref{fig:kohn}(b)]:
\begin{itemize}
\item $-\Pi_1/(N_Fr_s)$ rises from $\sqrt2/\pi=0.45$ at $q=0$ to $\approx1.21$ at $q=2k_F(1-10^{-4})$.
\item Above $2k_F$ it collapses precipitously, to 0.64 at $1.005\times2k_F$ and 0.25 at $1.05\times2k_F$.
\item The derivative is singular on both sides of $2k_F$, consistent with the singular logarithmic structure reported in Refs.~\cite{Maldague78,Czachor82,Khalil01}.
\end{itemize}
The Kohn anomaly is thus converted from a one-sided kink into a sharp, two-sided peak, whose weight grows linearly with $r_s$.

Two physical consequences follow. First, the $2k_F$ Fourier component of the screened potential is less screened, and the Friedel oscillations of the screening cloud are enhanced; Khalil \emph{et al.} found the same with screened interaction lines~\cite{Khalil01}. Second, $2k_F$ backscattering is sensitive to precisely this structure. For charged-impurity-limited transport in 2D semiconductor systems, the temperature dependence of the resistivity is controlled by the thermal smearing of the $2k_F$ kink in $\Pi(q)$~\cite{Stern80,GoldDolgopolov86,DasSarmaHwang99}. An enhanced, two-sided singularity would therefore strengthen those effects relative to RPA.

\subsection{Short wavelengths ($q>2k_F$)}
TF overscreens badly here, since the true $-\Pi/N_F$ falls as $2k_F^2/q^2$. The exchange correction also decays, and faster than $\Pi_0$: relative to the bubble it is $\approx0.13\,r_s$ at $q=2.5k_F$ and $\approx0.02\,r_s$ at $q=4k_F$. RPA is thus already adequate at large $q$, and the correction matters only in the immediate vicinity of $2k_F$.

\subsection{Verdict}
The calculated screening is a genuine, systematic improvement on TF theory in two respects:
\begin{enumerate}
\item The screening length is corrected at the exact-compressibility level.
\item The Fermi-surface ($2k_F$) structure, absent in TF and only weakly present in RPA, is qualitatively enhanced.
\end{enumerate}

The second improvement is, however, only qualitative. At $r_s=1$ the first-order correction near $2k_F$ exceeds the bare bubble itself [Fig.~\ref{fig:kohn}(a)], so perturbation theory is not controlled there. Moreover, in 3D the corresponding exchange peak has been argued to be largely cancelled by parallel-spin correlation~\cite{Kukkonen26}, whereas in 2D the fluctuation diagrams add further enhancement~\cite{Khalil01}. The sign of the effect, a strengthened Kohn anomaly, is robust at this order. Its magnitude at experimentally relevant $r_s$ requires correlation diagrams or QMC input. Naively resumming the $2k_F$ peak produces the spurious instability discussed in Sec.~\ref{sec:hubbard}.

\section{Friedel oscillations: RPA versus RPA + exchange}\label{sec:friedel}
\subsection{Setup}
The real-space consequence of the $2k_F$ structure is best seen in the density response to a point impurity of charge $Ze$ in the plane. With the RPA series of Eq.~(\ref{eq:series}),
\begin{gather}
\delta n(q)=\chi(q)V_{\rm ext}(q)=\frac{\Pi(q)\,V_{\rm ext}(q)}{1-v_q\Pi(q)},\qquad V_{\rm ext}=-Zv_q,\nonumber\\
\delta n(r)=\frac1{2\pi}\int_0^\infty\!dq\,q\,J_0(qr)\,\delta n(q).\label{eq:dnr}
\end{gather}
We compare $\Pi=\Pi_0$ (RPA) with $\Pi=\Pi_0+\Pi_1$ (RPA + exchange).

Both satisfy perfect screening, $\delta n(q\to0)=Z$. $\Pi_1$ was computed on a dense grid of 100 points. The grid is clustered geometrically toward $2k_F$, down to $|q-2k_F|=10^{-5}k_F$, and extends to $12k_F$. Beyond $12k_F$ we use the fitted power-law tail $P\propto q^{-5.0}$; this is immaterial.

Interpolation is monotone (PCHIP) in the variable $\sqrt{|q-2k_F|}$ on each side of $2k_F$. The Hankel transform uses composite Gauss--Legendre quadrature up to $q=200k_F$.

\subsection{RPA benchmark}
The asymptotic RPA result is fixed by the one-sided singularity $\Pi_0\simeq-N_F[1-\Theta(\delta)\sqrt{\delta/k_F}]$, where $\delta=q-2k_F$. Using $J_0(qr)\simeq\sqrt{2/\pi qr}\cos(qr-\pi/4)$ and $\int_0^\infty\delta^{1/2}e^{i\delta r}d\delta=\Gamma(\frac32)e^{3\pi i/4}r^{-3/2}$ gives
\begin{equation}
\frac{\delta n_{\rm RPA}(r)}{Zk_F^2}\simeq\frac{r_s/\sqrt2}{2\pi(1+r_s/\sqrt2)^2}\,\frac{\sin 2k_Fr}{(k_Fr)^2}+\frac{k_F}{2\pi q_{TF}(k_Fr)^3},\label{eq:FOrpa}
\end{equation}
where $1+r_s/\sqrt2=\varepsilon_{\rm RPA}(2k_F)$. The second term is the non-oscillatory tail from the $|q|$ nonanalyticity of $\delta n(q)=Z/(1+q/q_{TF}+\dots)$.

\begin{figure*}[t]
\centering
\includegraphics[width=0.95\textwidth]{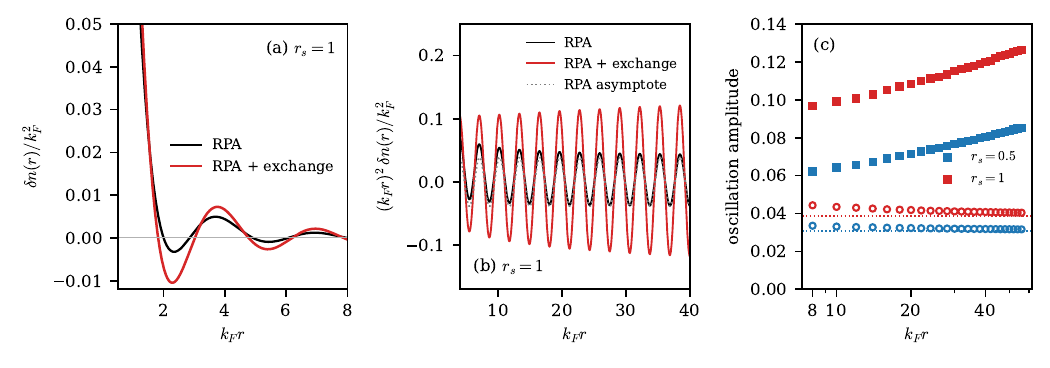}
\caption{Friedel oscillations around an in-plane point charge $Z=1$, from Eq.~(\ref{eq:dnr}). (a) Induced density $\delta n(r)/k_F^2$ at $r_s=1$: RPA (black) versus RPA + exchange, $\Pi=\Pi_0+\Pi_1$ (red). (b) $(k_Fr)^2\delta n(r)$ at $r_s=1$, compared with the analytic RPA asymptote of Eq.~(\ref{eq:FOrpa}) (dotted; oscillatory term only). The small upward offset of the numerical curves is the non-oscillatory $r^{-3}$ tail. (c) Oscillation amplitude from local fits of $(k_Fr)^2[\delta n-\delta n_{\rm smooth}]$ to $A\sin(2k_Fr+\varphi)$ over one period. Open symbols: RPA. Filled: RPA + exchange. Dotted lines: analytic RPA amplitude. The RPA amplitude saturates, whereas the exchange-corrected amplitude grows approximately as $\ln(k_Fr)$.}
\label{fig:friedel}
\end{figure*}

The numerical transform reproduces both terms. The fitted RPA oscillation amplitude approaches the analytic value, 0.0307 at $r_s=0.5$ and 0.0386 at $r_s=1$, with the expected $1/r$ corrections [Fig.~\ref{fig:friedel}(c)]. The phase is $\sin2k_Fr$ to within 0.05 rad. This confirms the accuracy of the procedure.

\subsection{Effect of exchange}
Three features stand out in Fig.~\ref{fig:friedel}:
\begin{enumerate}
\item \emph{Near the impurity} [panel (a)], the total screening charge is unchanged, but it is redistributed. The first density minimum at $k_Fr\approx2.3$ is three times deeper, and the subsequent oscillations are markedly larger.
\item \emph{Amplitude.} The oscillation amplitude is enhanced by a factor 1.9--2.7 at $r_s=0.5$, and 2.2--3.1 at $r_s=1$, over $8<k_Fr<56$. The phase remains essentially that of RPA ($\sin2k_Fr$, shifts $\lesssim0.03$ rad).
\item \emph{Growth with distance.} The RPA amplitude saturates, while the exchange-corrected amplitude keeps growing with distance, approximately as $\alpha+\beta\ln(k_Fr)$. We find $\beta\approx0.012$ at $r_s=0.5$ and $\beta\approx0.016$ at $r_s=1$ (in the units of Fig.~\ref{fig:friedel}), against RPA amplitudes of 0.031 and 0.040.
\end{enumerate}
The non-oscillatory tail is correspondingly reduced, to $k_F/(2\pi q_{TF}^*(k_Fr)^3)$.

\subsection{Origin of the logarithm}
The logarithm is the real-space image of the two-sided singularity of $\Pi_1$ at $2k_F$ (Sec.~\ref{sec:kohn}). On our dense grid, $P(q)$ is continuous at $2k_F$, with $P(2k_F)\approx-0.569$. It has slopes diverging on both sides, compatible with the $\sqrt{|\delta|}\ln|\delta|$ form proposed in Ref.~\cite{Khalil01}.

A term $\sqrt\delta\ln\delta$ transforms as
\begin{equation}
\int_0^\infty\!\delta^{1/2}\ln\delta\;e^{i\delta r}d\delta=\Gamma(\tfrac32)e^{3\pi i/4}r^{-3/2}\big[\psi(\tfrac32)+\tfrac{i\pi}{2}-\ln r\big],\label{eq:FTlog}
\end{equation}
so it produces precisely a $\ln(k_Fr)\,\sin(2k_Fr)/r^2$ tail. The contribution from $q<2k_F$, which is absent in RPA, adds further $r^{-2}$ terms. Our data resolve the singular structure well on the scale $|\delta|\gtrsim10^{-3}k_F$ that controls $k_Fr\lesssim10^2$. They cannot, however, discriminate reliably between $\sqrt{|\delta|}\ln|\delta|$ and admixtures such as $\sqrt{|\delta|}\ln^2|\delta|$. The $\ln r$ law in Fig.~\ref{fig:friedel}(c) should therefore be regarded as empirical over the range shown.

\subsection{Interpretation}
The exchange correction to the Friedel amplitude is thus of order $r_s\ln(k_Fr)$, a large logarithm. First-order perturbation theory fails for the \emph{asymptotic} tail once $r_s\ln(k_Fr)\sim1$, even at small $r_s$.

This is to be contrasted with the many-body local-field description of Giuliani and Simion~\cite{GiulianiSimion03,SimionGiuliani05}. There a static $G_+(q)$ that is smooth at $2k_F$ leaves the RPA form $\sin(2k_Fr)/r^2$ intact and only renormalizes its amplitude through $\varepsilon(2k_F)$. The exact first-order $\Pi_1$ is not of that form, because its own $2k_F$ singularity is stronger than that of $\Pi_0$. Equivalently, the exact $G_x$ of Fig.~\ref{fig:hub}(a) is itself singular at $2k_F$.

Whether higher orders exponentiate the logarithm, modifying the decay law as interactions do in one dimension, or cut it off, leaving an amplitude renormalization as assumed in local-field theory, cannot be decided at first order.

Robust conclusions at this order are:
\begin{itemize}
\item At intermediate distances ($k_Fr\lesssim50$), exchange enhances 2D Friedel oscillations by a factor of 2--3 at $r_s=0.5$--1, with unchanged period and phase.
\item RPA, and \emph{a fortiori} TF theory, which has no oscillations at all, substantially underestimates them.
\end{itemize}
This bears directly on STM imaging of impurities in 2D systems, on impurity-mediated effective interactions, and on the $2k_F$-dominated transport discussed in Sec.~\ref{sec:kohn}.

\subsection{Implications for Kohn--Luttinger pairing}
Kohn--Luttinger (KL) superconductivity~\cite{KohnLuttinger65} is the momentum-space counterpart of the Friedel oscillations. The nonanalyticity of the screened interaction $W(q)=v_q/\varepsilon(q)$ at $q=2k_F$ produces attractive components in the high angular-momentum channels of the pairing interaction on the Fermi surface. For pairing, only $q=|\bk-\bk'|=2k_F\sin(\theta/2)\le2k_F$ is sampled. Backscattering ($\theta\to\pi$) probes $q\to2k_F^-$.

In the 2D RPA, $\Pi_0$ is strictly constant for $q\le2k_F$. The Fermi-surface interaction $W(q)=2\pi e^2/\kappa(q+q_{TF})$ is then smooth, with no Kohn singularity on the relevant side. KL attraction is absent at lowest order and first appears only at higher order~\cite{Chubukov93}.

The present exchange correction changes this at $O(r_s)$. $\Pi_1$ is singular as $q\to2k_F^-$ [Fig.~\ref{fig:kohn}(b)], so $W(2k_F\sin\frac\theta2)$ acquires a nonanalytic cusp at $\theta=\pi$. Its angular harmonics then decay only as a power of $\ell$ with sign alternation, which generates attractive channels already at this order. Khalil \emph{et al.} reached a similar conclusion with screened interaction lines~\cite{Khalil01}.

The larger Friedel amplitude found above is the real-space expression of the same singularity. It therefore implies correspondingly larger KL couplings than RPA would suggest.

Three cautions apply:
\begin{enumerate}
\item The resulting pairing couplings arise from a singularity of relative weight $O(r_s)$ at the edge of the Fermi-surface integration, and remain small in the high-$\ell$ channels.
\item The growth of the Friedel amplitude with distance shows that first-order theory does not control the long-distance structure of the effective interaction, which is precisely what sets the KL pairing scale.
\item KL pairing relies on the sharpness of the $2k_F$ singularity, i.e., on Friedel oscillations persisting to large $r$. Both disorder (elastic mean free path $\ell_e$) and temperature (thermal length $v_F/T$) cut off the oscillations beyond $r\sim\min(\ell_e,v_F/T)$. This smears the singularity on the scale $|q-2k_F|\sim1/\min(\ell_e,v_F/T)$, and correspondingly suppresses the KL attraction in realistic samples, whatever its enhancement by exchange.
\end{enumerate}

\section{Comparison with the Hubbard local-field factor}\label{sec:hubbard}
\subsection{The two approximations}
In the Hubbard approximation (HA)~\cite{Hubbard57}, as adapted to 2D by Jonson~\cite{Jonson76}, the exchange hole is represented by replacing the exchange interaction $v_{\bk-\bk'}$ in the particle--hole ladder by $\frac12 v_{\sqrt{q^2+k_F^2}}$. The resulting ladder is then summed as $\Pi=\Pi_0/(1+v_qG_H\Pi_0)$, with
\begin{equation}
G_H(q)=\frac{q}{2\sqrt{q^2+k_F^2}}=\frac{x}{2\sqrt{1+x^2}}.\label{eq:GH}
\end{equation}
Like $G_x$, $G_H$ is independent of $r_s$. It is therefore to be judged as an approximation to the exact exchange-only local-field factor, i.e., to $G_x$ of Eq.~(\ref{eq:G}) and Fig.~\ref{fig:hub}(a). Expanded to first order in $e^2$, the HA corresponds to
\[
P_H(x)=-\frac{1}{\pi\sqrt{1+x^2}},
\]
to be compared with the exact $P(x)$ of Eq.~(\ref{eq:I}).

Three structural differences are evident:
\begin{enumerate}
\item The HA keeps only the vertex (ladder) channel. It omits the self-energy diagrams (b),(c) of Fig.~\ref{fig:diagrams}, and is therefore not a conserving approximation.
\item The HA replaces the singular kernel $v_{\bk-\bk'}$ by a smooth average. By Eqs.~(\ref{eq:PVas})--(\ref{eq:PSEas}), the true vertex and self-energy pieces are each $\propto\ln(k_F/q)$ and cancel only in combination. The HA's smoothing removes the divergence of the ladder by fiat, rather than through the cancellation enforced by the Ward identity.
\item $G_H$ is analytic in $q^2$ apart from the overall factor of $q$. It has no Fermi-surface structure, neither the $x^3\ln x$ term of Eq.~(\ref{eq:G}) nor any structure at $2k_F$.
\end{enumerate}

\begin{table}[b]
\caption{Exact first-order local-field factor $G_x=-(\pi/2)\,x\,P(x)$ versus the Hubbard form, Eq.~(\ref{eq:GH}).}\label{tab:hub}
\centering\small
\begin{tabular}{@{}lccc@{}}
\toprule
$x=q/k_F$ & $G_x$ & $G_H$ & $G_H/G_x$\\
\midrule
$\to0$ & $x/\pi$ & $x/2$ & $\pi/2=1.571$\\
0.2 & 0.06380 & 0.09806 & 1.537\\
0.5 & 0.16169 & 0.22361 & 1.383\\
1.0 & 0.34356 & 0.35355 & 1.029\\
1.5 & 0.59455 & 0.41603 & 0.700\\
1.9 & 1.03870 & 0.44245 & 0.426\\
1.98 & 1.34677 & 0.44631 & 0.331\\
\bottomrule
\end{tabular}
\end{table}

\subsection{Long wavelengths and the compressibility sum rule}
The compressibility sum rule in 2D reads~\cite{Davoudi01}
\[
G_+(q\to0)=A_+x,\qquad A_+=\frac{1-\kappa_0/\kappa}{\sqrt2\,r_s}.
\]
With the HF value $\kappa_0/\kappa=1-\sqrt2r_s/\pi$ this gives exactly $A_+=1/\pi$, the slope of $G_x$. The HA gives instead $A_+=1/2$, i.e., $\kappa_0/\kappa=1-r_s/\sqrt2$. This is wrong already at leading order in $r_s$, overestimating the exchange enhancement of the compressibility by the factor $\pi/2$. The analogous factor in 3D is 2 [$G_H\to x^2/2$ versus the exact $x^2/4$~\cite{GeldartTaylor70,EngelVosko90,Kukkonen26}].

The first-order result is also better than its formal order suggests. In 2D the correlation energy has the high-density form $\epsilon_c=a_0+b_0r_s\ln r_s+O(r_s)$~\cite{RajagopalKimball77,GoriGiorgi03}. Its constant term drops out of $\kappa_0/\kappa=1-\sqrt2r_s/\pi+\frac{r_s^4}{8}(\epsilon_c''-\epsilon_c'/r_s)$ (Rydberg units)~\cite{Davoudi01}. Correlation therefore enters $\kappa_0/\kappa$ only at $O(r_s^3\ln r_s)$: the exchange-only $\kappa_0/\kappa$ is exact through $O(r_s^2)$, and correspondingly $A_+=1/\pi+O(r_s^2\ln r_s)$.

Table~\ref{tab:comp} makes this quantitative. It uses the diffusion Monte Carlo (DMC) parametrization of $\epsilon_c$ from Refs.~\cite{Attaccalite02,GoriGiorgi03}:
\begin{itemize}
\item At $r_s=1$, $\kappa_0/\kappa=0.533$ (DMC), versus 0.550 from exchange only and 0.293 from the HA.
\item The DMC slope $A_+$ stays within 4\% of $1/\pi$ at $r_s=1$ and within 8\% at $r_s=2$, while the HA slope is 44--55\% too large.
\item The DMC compressibility diverges at $r_s\simeq2.03$. The resummed exchange-only form diverges at $\pi/\sqrt2=2.22$; the HA diverges at $\sqrt2=1.41$.
\end{itemize}

\begin{table}[t]
\caption{Inverse compressibility and small-$q$ slope $A_+$ of $G_+$. DMC values are obtained from the correlation-energy parametrization of Refs.~\cite{Attaccalite02,GoriGiorgi03} via the sum rule.}\label{tab:comp}
\centering\small
\begin{tabular}{@{}lcccccc@{}}
\toprule
 & \multicolumn{3}{c}{$\kappa_0/\kappa$} & \multicolumn{3}{c}{$A_+$}\\
\cmidrule(lr){2-4}\cmidrule(l){5-7}
$r_s$ & DMC & $G_x$ & HA & DMC & $G_x$ & HA\\
\midrule
0.5 & 0.772 & 0.775 & 0.646 & 0.323 & 0.318 & 0.5\\
1 & 0.533 & 0.550 & 0.293 & 0.330 & 0.318 & 0.5\\
2 & 0.018 & 0.100 & $-0.414$ & 0.347 & 0.318 & 0.5\\
3 & $-0.538$ & $-0.350$ & $-1.12$ & 0.363 & 0.318 & 0.5\\
\bottomrule
\end{tabular}
\end{table}

\subsection{Intermediate and large wave vectors}
\begin{figure*}[t]
\centering
\includegraphics[width=0.86\textwidth]{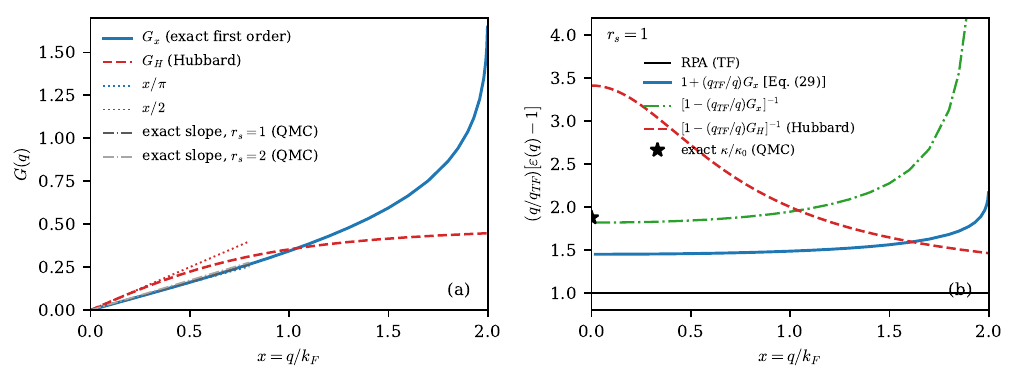}
\caption{(a) Exact first-order (exchange) local-field factor $G_x$ (solid), computed from $P(x)$, versus the Hubbard form $G_H$ (dashed). Dotted lines: small-$q$ asymptotes $x/\pi$ and $x/2$. Dash-dotted lines: the exact slopes $A_+$ at $r_s=1,2$ inferred from DMC compressibilities (Table~\ref{tab:comp}). (b) Screening enhancement $(q/q_{TF})[\varepsilon(q)-1]=-\Pi_{\rm eff}/N_F$ at $r_s=1$. Solid: the systematic $O(r_s)$ result, Eq.~(\ref{eq:eps}), i.e., the RPA series with $\Pi_0+\Pi_1$. Dash-dotted: the same $G_x$ resummed in Hubbard form. Dashed: Hubbard approximation. Star: exact $\kappa/\kappa_0$ from DMC.}
\label{fig:hub}
\end{figure*}

Figure~\ref{fig:hub}(a) and Table~\ref{tab:hub} compare the two local-field factors over $0<q<2k_F$:
\begin{itemize}
\item $G_H$ exceeds $G_x$ for $q\lesssim1.05k_F$, where the two cross.
\item $G_H$ saturates toward $1/2$, whereas $G_x$ rises steeply toward $2k_F$. It reaches $G_x\approx1.35$ at $q=1.98k_F$, three times $G_H$.
\end{itemize}
The steep rise reflects the singular (logarithmic) derivative of the first-order diagrams at $2k_F$ found by Maldague~\cite{Maldague78}, by Czachor \emph{et al.}~\cite{Czachor82}, and, with screened lines, by Khalil \emph{et al.}~\cite{Khalil01}. Our quadrature is not designed for $q\to2k_F$, and we do not quote $G_x(2k_F)$.

The HA's large-$q$ value $\frac12$ coincides with $1-g_x(0)$, the exchange-only value of the \emph{dynamic} high-frequency limit $G_+(\infty,\omega)=1-g(0)$~\cite{SantoroGiuliani88}. This is the one limit the HA reproduces by construction. The static first-order $G_x(q\to\infty)$ is a different limit (it is $1/3$ in 3D~\cite{GeldartTaylor70,Kukkonen26}), which we have not computed in 2D.

\subsection{Consequences for screening, and the resummation question}
Figure~\ref{fig:hub}(b) shows the screening enhancement at $r_s=1$. Three observations follow.

\emph{The HA at small $q$.} The HA's small-$q$ enhancement, 3.41, overshoots the exact $\kappa/\kappa_0=1.88$ by 80\%. The systematic first-order result gives 1.45; its Hubbard-type resummation gives 1.82.

\emph{The HA at intermediate $q$.} The HA's apparent reasonableness near $q\sim k_F$ is an accident of the crossing of $G_H$ and $G_x$.

\emph{The HA's ``sum of all ladders''.} Its resummation adds no controlled physics. Expanding it to $O(r_s)$ returns $G_H$ itself, which is already incorrect at that order. The geometric form merely promotes the $O(r_s)$ error into a spurious compressibility instability at $r_s=\sqrt2$.

\emph{Resumming the exact $G_x$ is not safe either.} It is instructive that the same resummation is not harmless with the exact $G_x$. Because $\sqrt2r_sG_x/x$ grows toward $2k_F$, $1-(q_{TF}/q)G_x$ vanishes near $2k_F$ once $r_s\gtrsim0.86$. This produces a spurious static $2k_F$ (charge-density-wave-like) instability, cf. the CDW discussion in Refs.~\cite{Maldague78,Czachor82}. In 3D the $2k_F$ enhancement of first-order exchange has been argued to be largely cancelled by parallel-spin correlation~\cite{Kukkonen26}, and 2D fluctuation diagrams are large near $2k_F$~\cite{Khalil01}. Neither the resummed $G_x$ nor its instability should therefore be trusted near $2k_F$ at finite $r_s$. The systematic object is the $O(r_s)$ dielectric function, Eq.~(\ref{eq:eps}), which is what the RPA series built on $\Pi_0+\Pi_1$ delivers.

\subsection{Which is better?}
On theoretical grounds the answer is unambiguous. The present calculation is a systematic, conserving (Ward-identity-respecting) many-body approximation: the self-energy and vertex diagrams are kept together. It is exact to $O(r_s)$ at every $q<2k_F$, and it satisfies the compressibility sum rule. In 2D the resulting long-wavelength compressibility $\kappa_0/\kappa$ is in fact exact through $O(r_s^2)$.

The Hubbard factor is an interpolation formula. It is not the leading term of any expansion, violates the sum rule at leading order, and misses the nonanalytic small-$q$ and $2k_F$ structure.

Quantitatively, $G_x$ is also the better long-wavelength local-field factor through $r_s\approx2$ (Table~\ref{tab:comp}). The HA's only advantages are its closed form and a sensible saturation at large $q$.

Neither approximation contains correlation. For $r_s\gtrsim1$ near $2k_F$, where fluctuation (two-interaction-line) diagrams become comparable to the exchange terms~\cite{Khalil01}, a QMC-constrained $G_+$~\cite{Davoudi01} is required. The present $G_x$ provides the exact high-density limit that any such parametrization must reproduce.

\section{Relation to previous work}\label{sec:prior}
\subsection{Earlier computations of the same diagrams}
The first-order (exchange) correction $\Pi_1(q)$ to the static polarizability of the strictly 2D electron gas with bare Coulomb interaction has been computed before:
\begin{itemize}
\item Maldague computed it numerically in 1978, in connection with charge-density-wave instabilities in layered compounds~\cite{Maldague78}.
\item Czachor, Holas, Sharma, and Singwi evaluated all first-order diagrams for arbitrary $q$ and $\omega$, examined their singularities and analytic properties, and noted the strong static enhancement near $2k_F$~\cite{Czachor82}.
\item Their perturbative $G$ served as the reference for the variational dynamical-exchange-decoupling local-field factor of Hameeuw, Brosens, and Devreese~\cite{Hameeuw03}.
\item Khalil, Ashcroft, and Teter recomputed the self-energy and exchange diagrams with dynamically screened interactions~\cite{Khalil01}. They added the fluctuation diagrams, characterized the $2k_F$ singularity, and quoted the $q\to0$ enhancement $0.45\,r_s$ ($=\sqrt2r_s/\pi$).
\end{itemize}
In 3D, the corresponding exact static exchange result goes back to Geldart and Taylor~\cite{GeldartTaylor70} and Engel and Vosko~\cite{EngelVosko90}. It was recently reformulated as an exact exchange-only local-field benchmark by Kukkonen~\cite{Kukkonen26}.

\subsection{Present contribution}
The earlier 2D calculations~\cite{Maldague78,Czachor82,Khalil01} are numerical, and the $q\to0$ limit itself is fixed by the compressibility sum rule~\cite{Davoudi01}. To our knowledge, the following analytic results are new:
\begin{enumerate}
\item the closed-form logarithmic asymptotics of the separate self-energy and vertex pieces, Eqs.~(\ref{eq:PVas})--(\ref{eq:PSEas}), including the constants $4\ln2+3/2$ and $4\ln2-1/2$;
\item the proof that the $O(q)$ term vanishes;
\item the $q^2\ln q$ structure with analytic coefficient $1/48\pi^2$ and the constant $\beta$, and hence the $x^3\ln x$ term in $G_x$;
\item the uniform-lune representation, Eqs.~(\ref{eq:jac})--(\ref{eq:I});
\item the quantitative, sum-rule-based comparison with the Hubbard factor, including the $O(r_s^3\ln r_s)$ accuracy of the exchange-only compressibility.
\end{enumerate}

\section{Conclusion}
In this work we calculated the leading corrections to the 2D RPA dielectric screening in the static limit in the conserving approximation, including the first-order self-energy and vertex corrections to the bare irreducible polarizability function. The importance of conserving approximations for the dielectric function, associated with the Ward identities, has been known for a long time~\cite{DasSarmaHanke83}, and \emph{ad hoc} approximations incorporating local-field corrections, such as the Hubbard approximation~\cite{Hubbard57,Jonson76} and various self-consistent screening approximations~\cite{STLS68}, manifestly violate the Ward identities and should be avoided.

\section*{Acknowledgments}
This work is supported by the Laboratory for Physical Sciences. The work was assisted by Claude Opus AI in helping with numerics and producing figures.

\begin{center}\small\bfseries APPENDIX\end{center}
\appendix
\section{Exchange self-energy in 2D}\label{app:sigma}
With $E$ and $K$ the complete elliptic integrals of modulus $k$,
\begin{equation}
\Sigma_x(k)=-\frac{2e^2}{\pi\kappa}\begin{cases}k_F\,E(k/k_F), & k<k_F,\\[2pt]
k\big[E(\tfrac{k_F}{k})-(1-\tfrac{k_F^2}{k^2})K(\tfrac{k_F}{k})\big], & k>k_F.\end{cases}\label{eq:SigA}
\end{equation}
This form was verified against direct integration of Eq.~(\ref{eq:Sigx}) to $10^{-12}$. Using $dE/dk=(E-K)/k$ and $K\simeq\ln(4/k')$ gives $\partial_k\Sigma_x\simeq(e^2/\pi\kappa)[\ln(8k_F/|k-k_F|)-2]$ near $k_F$.

\section{Derivation of Eq.~(\ref{eq:jac})}
For $p_x>0$, the inner boundary is $|\bp+\bq/2|=k_F$ and the outer boundary is $|\bp-\bq/2|=k_F$. Their radii at angle $\theta$ are the roots of $r^2\pm rq\cos\theta+q^2/4=k_F^2$, namely $r_\mp=\rho(\theta)\mp\frac q2\cos\theta$. The region $p_x<0$ follows by mirror symmetry. The width $r_+-r_-=q|\cos\theta|$ exactly cancels the $1/|p_x|$ of $L_{\bk}$. This is why the Lindhard bubble is exactly $-N_F$ for $q<2k_F$, and why every $1/D$ in Eq.~(\ref{eq:kernel}) becomes a uniform measure.

\end{document}